\documentclass[11pt]{article} 
\usepackage[T1]{fontenc}           
\usepackage{lmodern}               
\usepackage{microtype}             

\usepackage[margin=1in]{geometry} 

\usepackage{amsmath,amssymb,amsthm}
\usepackage{bm}                    
\usepackage{graphicx}
\usepackage{hyperref}
\usepackage{booktabs}  
\usepackage{placeins}  
\usepackage{fancyvrb}  
\usepackage{booktabs}              

\usepackage{natbib} 
\usepackage{listings}
\usepackage{xcolor}

\definecolor{codeblue}{rgb}{0,0,0.6}
\definecolor{codegreen}{rgb}{0,0.5,0}
\definecolor{codered}{rgb}{0.6,0,0}

\lstdefinestyle{academic}{
    commentstyle=\color{codegreen},
    keywordstyle=\color{codeblue},
    stringstyle=\color{codered},
    basicstyle=\ttfamily\small,
    breaklines=true,                 
    captionpos=b,                    
    numbers=none,
    tabsize=2,
    showstringspaces=false
}
\lstdefinelanguage{Solidity}{
  keywords={contract, function, mapping, public, view, returns, bool, address, uint256, internal, constant, override, if},
  keywordstyle=\color{codeblue},
  ndkeywords={msg, sender},
  ndkeywordstyle=\color{codeblue},
  comment=[l]{//},
  morecomment=[s]{/*}{*/},
  commentstyle=\color{codegreen},
  stringstyle=\color{codered},
  morestring=[b]',
  morestring=[b]",
  sensitive=true
}

\title{Transaction Proximity:\\%
       A Graph-Based Approach to Blockchain Fraud Prevention}
\author{%
  Gordon Liao%
  \and Ziming Zeng%
  \and Mira Belenkiy%
  \and Jacob Hirshman%
}
\date{May 2025}

\begin{document}

\begin{titlepage}

    \maketitle
    \begin{abstract}
        This paper introduces a fraud-deterrent access validation system for public blockchains, leveraging two complementary concepts: "Transaction Proximity", which measures the distance between wallets in the transaction graph, and "Easily Attainable Identities (EAIs)", wallets with direct transaction connections to centralized exchanges. Recognizing the limitations of traditional approaches like blocklisting (reactive, slow) and strict allow listing (privacy-invasive, adoption barriers), we propose a system that analyzes transaction patterns to identify wallets with close connections to centralized exchanges. 
        
        Our directed graph analysis of the Ethereum blockchain reveals that 56\% of large USDC wallets (with a lifetime maximum balance greater than \$10,000) are EAI and 88\% are within one transaction hop of an EAI. For transactions exceeding \$2,000, 91\% involve at least one EAI. Crucially, an analysis of past exploits shows that 83\% of the known exploiter addresses are not EAIs, with 21\% being more than five hops away from any regulated exchange. We present three implementation approaches with varying gas cost and privacy tradeoffs, demonstrating that EAI-based access control can potentially prevent most of these incidents while preserving blockchain openness. Importantly, our approach does not restrict access or share personally identifiable information, but it provides information for protocols to implement their own validation or risk scoring systems based on specific needs. This middle-ground solution enables programmatic compliance while maintaining the core values of open blockchain.
    \end{abstract}
    
    \vfill
    
    \medskip

    \noindent\textbf{Acknowledgments:} We thank Joanna Marathakis and Caroline Hill for their helpful comments. We also thank Chris Brummer for valuable early discussions.
\medskip
    
    \noindent\textit{Note: The views expressed in this paper are solely those of the authors and do not necessarily reflect those of Circle Internet Financial or any other affiliated organizations.}

\end{titlepage}

\setcounter{page}{1}

\section{Introduction}

Public blockchains are revolutionizing finance by enabling trustless transactions without intermediaries. However, this openness comes with significant challenges, particularly with respect to fraud and exploits that have resulted in billions of dollars in losses\citep{chainalysis_crypto_crime_2024}. These security breaches erode user trust and hinder the widespread adoption of decentralized finance (DeFi) applications. Traditional mitigation approaches face fundamental limitations: blocklisting is reactive and often ineffective, requiring lengthy legal procedures for asset freezing after exploits have occurred, while strict allow listing based on Know Your Customer (KYC) standards faces significant resistance due to privacy concerns and adoption barriers within the DeFi ecosystem.

\begin{figure}[h]
    \centering
    \includegraphics[width=0.7\textwidth]{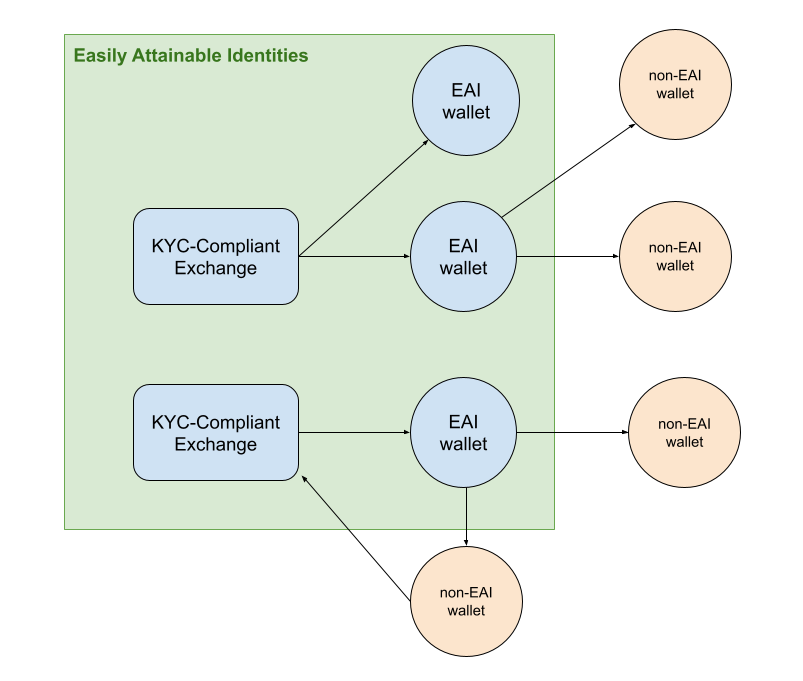}
    \caption{Transaction Proximity and Easily Attainable Identities (EAIs). This diagram illustrates the relationship between centralized exchanges and wallet addresses in the transaction graph, highlighting how transaction proximity can be used to identify EAIs.}
    \label{fig:eaidiagram}
\end{figure}

This paper introduces a middle-ground approach that combines two complementary concepts: Transaction Proximity and Easily Attainable Identities (EAIs). Transaction Proximity refers to the distance between wallets in the transaction graph, providing a quantitative measure of how closely connected different addresses are based on their transaction history. Easily Attainable Identities (EAIs) are wallet addresses within one transaction hop of centralized exchanges, which are increasingly becoming more regulated and centralized. As such, EAIs represent addresses for which identifying information could potentially be obtained from centralized exchanges if necessary for fraud and exploit investigation. Figure \ref{fig:eaidiagram} illustrates this assignment of EAIs within a transaction graph.

Our approach is built on a key insight: a significant portion of legitimate blockchain activity occurs within a few transaction hops of centralized exchanges. This is in part because omnibus wallets at these exchanges help preserve user privacy by breaking up transactions that otherwise would be linked together. By analyzing the transaction graph and identifying wallet addresses with close proximity to regulated entities, we can improve security without imposing universal KYC requirements or sharing personal identifying information (PII). This approach aligns with emerging research on privacy-preserving compliance mechanisms that seek to balance regulatory requirements with user privacy through cryptographic and algorithmic solutions\citep{duffie2025privacy}.

It is crucial to emphasize that our approach does not restrict access to blockchain protocols or impose universal requirements on users. Rather, it extracts and propagates readily available blockchain data that protocols and exchanges can use to implement their own control or risk mitigation based on their specific needs and risk tolerance. This flexibility allows each protocol to determine how to use transaction proximity and EAI data - whether for enhanced monitoring, adjusting collateral requirements, or implementing tiered access to certain features - without compromising the open nature of public blockchains.


Our approach is particularly relevant in the context of ongoing regulatory discussions. The traditional finance world relies heavily on Anti-Money Laundering and Combating the Financing of Terrorism (AML/CFT) regulations to combat illicit activities. In the United States, the Bank Secrecy Act (BSA) forms the backbone of this framework, mandating KYC procedures, transaction reporting, and record keeping for financial institutions. However, these regulations are outside the scope of decentralized finance, where transactions are facilitated most often by software without the involvement of financial intermediaries.\footnote{In particular, many of the common protocols used in decentralized finance are immutable software code that operates autonomously independent of any specific entity.} We contend that the public policy goals of the BSA can be achieved through alternative means that better align with the architecture and principles of blockchain technology.

Our empirical analysis demonstrates that the vast majority of transactions in stablecoins are carried out by wallets with close proximity to centralized exchanges. This finding opens opportunities for programmatic compliance approaches that can achieve regulatory objectives without sacrificing the fundamental benefits of open blockchains.
\medskip

Beyond fraud prevention, our transaction proximity and EAI framework enables numerous applications across the DeFi ecosystem:

\begin{itemize}
    \item \textbf{Risk-Based Lending:} Lending protocols can adjust collateral requirements based on a borrower's transaction proximity, offering more favorable terms to EAIs while maintaining higher requirements for those with a greater distance from regulated entities.
    
    \item \textbf{Reputation Systems:} Transaction proximity can serve as a foundation for onchain reputation systems that do not require direct identity disclosure, enabling protocols to establish graduated trust levels based on transaction graph positioning.
    
    \item \textbf{Selective Disclosure:} Projects can implement tiered access to features or services based on transaction proximity, creating incentives for legitimate use without mandating full KYC.
    
    \item \textbf{Governance Participation:} DAOs could implement weighted voting mechanisms that consider transaction proximity as a factor in determining voting power, potentially reducing the impact of Sybil attacks.
\end{itemize}


\textbf{Related Work} The systematic analysis of blockchain transaction patterns for security and compliance purposes was pioneered by Ron and Shamir's foundational work on Bitcoin transaction graphs~\citep{ron2013quantitative}. This foundational approach to transaction graph analysis has been extended by modern fraud detection systems employing machine learning techniques~\citep{taher2024fraud}. Building on these graph analysis foundations, our Transaction Proximity approach extends Ron and Shamir's entity clustering methodology by specifically focusing on proximity to regulated exchanges rather than general entity identification.

Parallel developments in privacy-preserving compliance have explored methods to balance user privacy with regulatory requirements. Duffie, Olowookere, and Veneris propose compliance-by-design stablecoin systems that embed privacy-preserving mechanisms directly into distributed ledgers using zero-knowledge proofs for KYC verification~\citep{duffie2025privacy}. Similarly, Gross et al. demonstrate how zero-knowledge proofs can enable cash-like private transactions within regulatory limits~\citep{gross2022compliant}, while Buterin et al. introduce Privacy Pools that allow users to prove membership in or exclusion from specific association sets without revealing complete transaction histories~\citep{buterin2024privacy}. The Privacy Pools framework particularly resonates with our approach, as both systems create separating equilibria between compliant and non-compliant users—their analysis showing honest users naturally excluding malicious actors from association sets mirrors our empirical finding that 83\% of known exploiter addresses are not EAIs.

A complementary strand of research focuses on adapting regulatory frameworks to leverage blockchain technology's unique features rather than forcing existing structures onto incompatible architectures. Brummer demonstrates this approach in the context of disclosure requirements, showing how traditional 1930s securities law disclosure frameworks are poorly suited to DeFi applications and proposing blockchain-native solutions like "Disclosure NFTs" and "Disclosure DAOs" to modernize regulatory compliance~\citep{brummer2022disclosure}. This philosophy of technological adaptation aligns with our transaction proximity framework, which similarly leverages existing blockchain infrastructure to achieve regulatory objectives.


Our transaction proximity and EAI framework contributes to this evolving landscape by providing a middle-ground approach that leverages existing blockchain infrastructure while enabling compliance verification without requiring universal KYC or complex cryptographic protocols and computational costs. Unlike approaches that require active user participation, our method operates passively on existing blockchain data, extending the foundational graph analysis techniques while addressing the growing need for scalable compliance solutions in decentralized systems.

The remainder of this paper is organized as follows. Section 2 details our methodology for implementing transaction proximity analysis and identifying EAIs through directed graph analysis of blockchain transactions. Section 3 presents empirical results on the prevalence of EAIs in different wallet sizes and transaction values, as well as evidence of their effectiveness in deterring exploits. Section 4 explores the implementation options for EAI-based access control systems, including onchain registries, off-chain solutions, Merkle trees, and hybrid approaches, with a comparative analysis of their performance and tradeoff characteristics. Finally, Section 5 concludes with implications for the future of blockchain security, trust establishment, and regulatory compliance.

\FloatBarrier

\section{Methodology}

This section outlines our approach for quantifying transaction proximity between wallet addresses and identifying Easily Attainable Identities (EAIs), as well as measuring the EAI characteristics of transactions.

\subsection{Graph Construction and EAI Identification}

We model the Ethereum blockchain as a directed graph to systematically evaluate the proximity of wallets to centralized exchanges. In this graph representation:

\begin{align}
G &= (V, E) \\
V &= \{v_1, v_2, \ldots, v_n\} \text{ (wallet addresses)} \\
E &\subseteq V \times V \text{ (directed transaction edges)}
\end{align}
where each node represents an individual wallet address and directed edges represent transactions between wallets. An edge $(v_i, v_j) \in E$ exists if wallet $v_i$ has transferred at least \$10 worth of ETH or major stablecoins (USDC/USDT) to wallet $v_j$:

\begin{equation}
(v_i, v_j) \in E \iff \text{total\_transfer}(v_i \to v_j) \geq \$10
\end{equation}

We specifically use a directed rather than bidirectional graph because users have control over where they send funds but not over who sends funds to them. This distinction is crucial to accurately assess the attainment of identity.

Let $\mathcal{EX} = \{e_1, e_2, \ldots, e_k\}$ represent the set of verified centralized exchange wallet addresses. We define the distance of any wallet $v$ to centralized exchange wallets as:

\begin{equation}
d(v) = \min_{e \in \mathcal{EX}} \{\text{shortest\_path\_length}(v, e)\}
\end{equation}

where $\text{shortest\_path\_length}(v, e)$ denotes the length of the shortest directed path from wallet $v$ to exchange wallet $e$ in graph $G$.

Easily Attainable Identities (EAIs) are then formally defined as:

\begin{equation}
\text{EAI}(v) = \begin{cases}
1 & \text{if } d(v) \leq 1 \\
0 & \text{otherwise}
\end{cases}
\end{equation}

This classification includes exchange wallet addresses themselves ($d(v) = 0$) and wallet addresses that have directly received funds from these exchange wallets ($d(v) = 1$).

\subsection{Distance Calculation Algorithm}

Our analytical process consists of four key steps:

\begin{enumerate}
    \item \textbf{Data Collection:} We compile a comprehensive list of verified exchange addresses and extract complete transaction data from the Ethereum blockchain.
    \item \textbf{Graph Construction:} We build the directed graph $G$ as defined above.
    \item \textbf{Breadth-First Search (BFS):} We implement BFS algorithms starting from all exchange addresses simultaneously to compute $d(v)$ for each wallet $v$.
    \item \textbf{EAI Classification:} We apply Equation (5) to classify wallets as EAIs.
\end{enumerate}

The BFS algorithm has computational complexity $O(|V| + |E|)$ and efficiently computes distances for all wallets in a single traversal. Due to computational constraints, we limit our BFS traversal to a maximum of 5 hops, which captures 98.2\% of USDC holders with balances of at least \$10.

\subsection{Transaction Classification}

Beyond wallet classification, we analyze the transaction proximity and EAI characteristics of individual transactions. For each transaction $T$ with sender $s$ and receiver $r$, we define the transaction's EAI distance as:

\begin{equation}
d_T = \min(d(s), d(r))
\end{equation}

This approach allows us to quantify the degree to which a transaction is connected to identifiable entities. The transaction is classified as involving an EAI if:

\begin{equation}
\text{EAI}_T = \max(\text{EAI}(s), \text{EAI}(r))
\end{equation}

meaning the transaction involves an EAI if either the sender or receiver is classified as an EAI.

\subsection{Limitations and Scope}

Our analysis covers the entire Ethereum blockchain from inception until May 31, 2024, resulting in a graph with 206 million nodes and 442 million edges. We acknowledge several important limitations:


\begin{itemize}
    \item \textbf{Incomplete Exchange Coverage}: Our analysis may not include all exchange wallets, which likely underestimate the true prevalence of EAIs in the ecosystem. The exchange addresses were collected from public sources and therefore may not be fully accurate. 
    \item \textbf{False Positives/Negatives}: Proximity to exchanges does not guarantee legitimate activity, and conversely, some legitimate users may operate at greater distances from exchanges.\footnote{For instance, exchange hacks can result in sending of funds from exchange addresses directly to exploiter addresses. These exchange hack addresses should be excluded from the EAI assignment.} As such, it would be important for exchanges to monitor transaction proximity and patterns to mitigate risks. 
    \item \textbf{Privacy Considerations}: Although our method focuses on identity attainability rather than direct identification, the transaction tracing approach still raises privacy considerations.
    \item \textbf{Exchange KYC Effectiveness}. The attainability of identity through transaction tracing can depend heavily on the historical and ongoing effectiveness of exchange-level KYC and monitoring practices, which may vary significantly across platforms and time.

\end{itemize}

\section{Empirical Results}

Our analysis of the Ethereum blockchain reveals several key insights about the prevalence and distribution of Transaction Proximity and Easily Attainable Identities (EAIs) across different wallet types and transaction patterns. These findings demonstrate the potential effectiveness of EAI-based approaches for fraud deterrence.

\subsection{Prevalence of EAIs}

We focused our analysis on USDC, a widely-used stablecoin, as a representative case study. This approach provides a manageable scope, while still offering meaningful insights that could be extended to other tokens in future implementations.

Our examination of the Ethereum blockchain (as of May 31, 2024) identified 1,226,629 externally owned accounts (EOAs) with maximum lifetime USDC balances exceeding \$10,000 USD. Among these significant wallets:

\begin{itemize}
    \item 687,444 wallets (56.0\%) are directly classified as EAIs
    \item 1,083,557 wallets (88.3\%) are within just one transaction hop of an EAI
\end{itemize}

\begin{figure}[ht]
    \centering
    \includegraphics[width=0.85\textwidth]{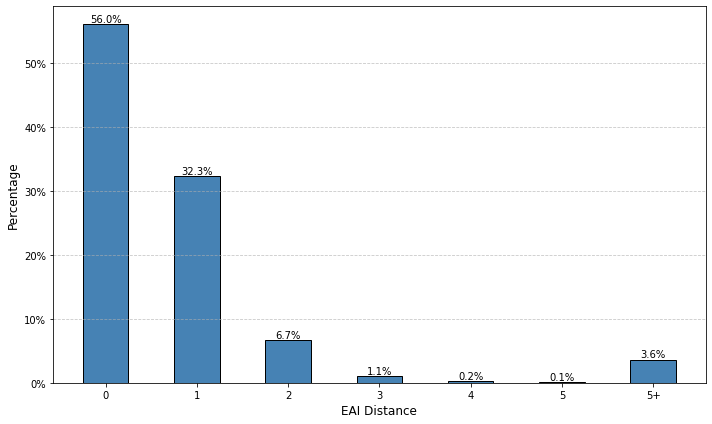}
    \caption{Distance to EAIs for USDC wallets with maximum lifetime balances greater than \$10,000. The graph shows a clear concentration at shorter distances, with 56.0\% of wallets being EAIs (distance 0) and 88.3\% within one transaction hop of an EAI.}
    \label{fig:distance}
\end{figure}

We selected the \$10,000 threshold to align with cash transaction reporting requirements under the Bank Secrecy Act (BSA), providing a relevant benchmark for regulatory considerations. Figure \ref{fig:distance} illustrates the distribution of wallet distances to EAIs, showing a clear concentration at shorter distances.

\FloatBarrier

\subsection{EAI Distance Consistency Across Wallet Sizes}

An important question is whether EAI distance varies significantly based on wallet balance size. As shown in Table \ref{tab:Distance_by_size}, the distribution of EAI distances remains relatively consistent across different balance categories. This consistency suggests that the relationship between wallets and identifiable entities is a fundamental characteristic of the ecosystem rather than being dependent on wallet size.

\begin{table}[h]
    \centering
    \caption{Number of Ethereum USDC EOA addresses by EAI distance and maximum lifetime balance as of 2024-05-31}\label{tab:Distance_by_size}
    \begin{tabular}{lrrrrrrr}
        \toprule
        & \multicolumn{7}{c}{\textbf{EAI distance}} \\
        \cmidrule(lr){2-8}
        \textbf{USDC Balance} & \textbf{0} & \textbf{1} & \textbf{2} & \textbf{3} & \textbf{4} & \textbf{5} & \textbf{5+} \\
        \midrule
        10-1k & 1,819,036 & 1,624,685 & 611,890 & 101,905 & 16,500 & 4,332 & 671,986 \\
        1k-100k & 1,676,942 & 1,082,669 & 251,404 & 33,715 & 5,396 & 1,389 & 229,270 \\
        100k-10m & 162,689 & 101,983 & 20,271 & 4,872 & 1,136 & 194 & 8,628 \\
        10m+ & 4,813 & 2,418 & 869 & 99 & 17 & 6 & 218 \\
        \bottomrule
    \end{tabular}
\end{table}

\subsection{Traceability of Transactions}

\begin{figure}[ht]
    \centering
    \includegraphics[width=0.85\textwidth]{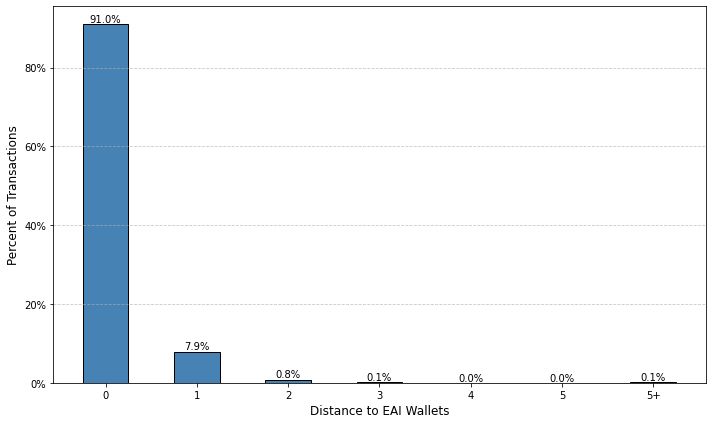}
    \caption{Distance to EAIs for USDC wallet-to-wallet transactions greater than \$2,000.}
    \label{fig:distancetxn}
\end{figure}

Moving beyond static wallet balances, we analyzed the transaction proximity and EAI characteristics of actual transactions. Figure \ref{fig:distancetxn} reveals that large USDC transfers are highly traceable through the transaction proximity framework. For USDC wallet-to-wallet transfers exceeding \$2,000, 91\% involve at least one wallet that is an EAI. This high percentage indicates that most significant transactions occur within a network where identifying information is potentially obtainable if needed for fraud investigation.

This pattern of high traceability holds consistently across different transaction value ranges. Table \ref{tab:kyc-distance-usdc-transaction-count} shows the distribution of transaction counts by EAI distance for various transfer amounts, from small (\$10-\$2,000) to very large (over \$10 million). The proportion of transactions at each EAI distance remains relatively stable across these categories, suggesting that the transaction proximity framework provides consistent traceability regardless of transaction size.

\begin{table}[h]
    \centering
    \caption{Number of Ethereum USDC wallet-to-wallet transfers by EAI distance and transfer amount from inception to 2024/05/31}
    \label{tab:kyc-distance-usdc-transaction-count}
    \resizebox{\textwidth}{!}{%
    \begin{tabular}{lrrrrrrr}
        \toprule
        & \multicolumn{7}{c}{\textbf{EAI distance}} \\
        \cmidrule(lr){2-8}
        \textbf{USDC Transfer Value} & \textbf{0} & \textbf{1} & \textbf{2} & \textbf{3} & \textbf{4} & \textbf{5} & \textbf{5+} \\
        \midrule
        10-2k & 16,876,678 & 2,334,333 & 317,150 & 39,870 & 8,597 & 1,584 & 41,461 \\
        2k-100k & 10,622,657 & 998,286 & 102,090 & 12,698 & 2,094 & 607 & 16,799 \\
        100k-10m & 2,187,091 & 127,284 & 15,838 & 2,767 & 595 & 111 & 1,747 \\
        10m+ & 66,764 & 3,011 & 341 & 29 & 10 & 2 & 11 \\
        \bottomrule
    \end{tabular}%
    }
\end{table}

\FloatBarrier

When examining the total value of transactions (Table \ref{tab:kyc-distance-usdc-transaction-volume}), we observe a similar pattern. The vast majority of transaction value occurs at EAI distances of 0 or 1, further confirming that most economic activity on the blockchain happens in close proximity to identifiable entities.

\begin{table}[h]
    \centering
    \caption{USD transfer value (millions) of Ethereum USDC wallet-to-wallet transfers by EAI distance and transfer amount from inception to 2024/05/31 }
    \label{tab:kyc-distance-usdc-transaction-volume}
    \begin{tabular}{lrrrrrrr}
        \toprule
        & \multicolumn{7}{c}{\textbf{EAI distance}} \\
        \cmidrule(lr){2-8}
        \textbf{USDC Transfer Value} & \textbf{0} & \textbf{1} & \textbf{2} & \textbf{3} & \textbf{4} & \textbf{5} & \textbf{5+} \\
        \midrule
        10-2k & 8,931 & 1,039 & 123 & 14 & 3 & 1 & 17 \\
        2k-100k & 164,112 & 13,463 & 1,332 & 175 & 36 & 12 & 265 \\
        100k-10m & 1,912,799 & 97,016 & 13,684 & 1,767 & 497 & 55 & 871 \\
        10m+ & 2,250,922 & 72,147 & 6,736 & 503 & 141 & 36 & 244 \\
        \bottomrule
    \end{tabular}
\end{table}

These findings collectively demonstrate that the vast majority of significant USDC transactions on Ethereum involve at least one party with a transaction proximity validation, providing a strong foundation for fraud-deterrent mechanisms based on EAI status.

\subsection{Deterrence Potential: Analysis of Known Exploiter Addresses}

\begin{figure}[htbp]
    \centering
    \includegraphics[width=0.85\textwidth]{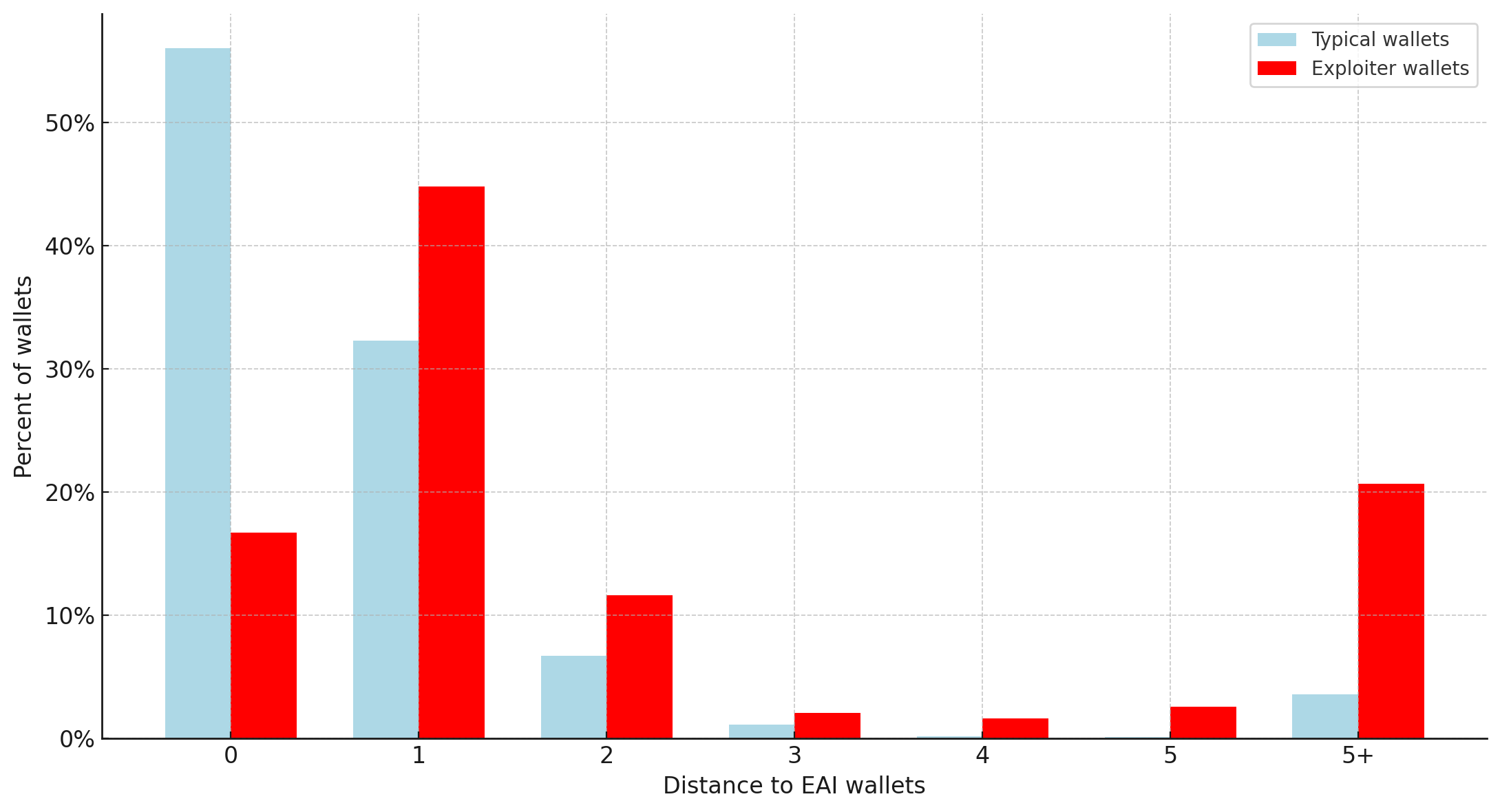}
    \caption{Distance to EAIs for Known Exploiter Addresses versus Typical Wallets. The distribution of exploiter wallets (in red) differs significantly from the general wallet sample (in blue), with 83\% of exploiter addresses being non-EAIs and 21\% being more than 5 hops away from any EAI. }
    \label{fig:hacker_dist}
\end{figure}

We evaluated EAI-based access controls as a deterrent against malicious activity by analyzing 431 recent exploiter wallet addresses from the post-May 2022 period, when large exchanges began mandating KYC or substantially increasing KYC requirements. These addresses, identified via Dune Analytics public labels, show dramatically different EAI distance patterns compared to typical wallets, as illustrated in Figure \ref{fig:hacker_dist}.

Our analysis reveals that:
\begin{itemize}
    \item 83\% of known exploiter addresses are not EAIs
    \item 21\% of exploiter addresses are more than 5 hops away from any EAI
\end{itemize}

This exploiter wallet distance-to-EAI distribution differs significantly from the general wallet sample, where the majority of addresses are within 0-1 hops of an EAI. The clear separation in distribution highlights the transaction pattern differences between legitimate users (who tend to operate close to EAIs) and malicious actors (who tend to operate at greater distances) provides compelling evidence for the potential effectiveness of EAI-based access controls as a fraud deterrence mechanism.

\FloatBarrier

\section{Implementation Approaches}

Based on our empirical findings, we now explore practical approaches for implementing a EAI-based wallet identification. The core technical challenge involves efficiently storing, maintaining, and verifying a list of EAI addresses within blockchain smart contracts. Note that these implementation approaches do not expose any additional information that is not already visible on public blockchain ledgers. Rather, they are aimed at making EAI tagging more real-time and easily queried for onchain applications. 

We evaluate implementation strategies, each with different tradeoffs regarding gas efficiency, transparency, and decentralization. These approaches represent a spectrum of options that can be tailored to specific use cases and security requirements.


All implementation approaches share a common goal: creating and maintaining a registry of EAI addresses that can be efficiently queried during transaction processing. This registry serves as the foundation for access control mechanisms that can be integrated into smart contracts, tokens, or decentralized applications.

The three primary implementation strategies we evaluated are the following:

\begin{enumerate}
    \item \textbf{On-Chain Registry:} Storing the complete list of EAI addresses directly onchain
    \item \textbf{Off-Chain Registry with On-Chain Verification:} Maintaining the registry off-chain while using cryptographic signatures for onchain verification
    \item \textbf{On-Chain Merkle Trees:} Representing the entire registry as a single Merkle root hash onchain, with proofs provided during transactions
 
\end{enumerate}

Each approach involves fundamental tradeoffs between initial deployment costs, per-transaction costs, transparency, and maintenance complexity. We discuss these tradeoffs in detail below and present performance benchmarks to guide implementation decisions.

\subsubsection*{On-Chain Registry}

The most straightforward implementation stores the complete EAI address registry directly on the blockchain. The registry only stores the list of pseudonymous addresses that are EAIs without compromising user data. This approach offers several advantages:

\begin{itemize}
    \item \textbf{Simplicity:} The implementation is straightforward, requiring only a mapping from addresses to boolean values
    \item \textbf{Transaction Efficiency:} Checking if an address is in the registry requires only a single storage read operation (O(1) complexity)
    \item \textbf{Transparency:} The pseudonymous address list of EAI wallets exists onchain and can be audited by anyone
    \item \textbf{Immediacy:} Updates to the registry can be programmed to be immediate following token transfer
\end{itemize}

A basic implementation would look like this:

\begin{lstlisting}[language=Solidity]
    mapping(address => bool) public eaiRegistry;
    
    function isEAI(address account) public view returns (bool) {
        return eaiRegistry[account];
    }
\end{lstlisting}

However, this approach has a significant drawback: the initial cost of populating the registry is extremely high. With millions of EAI addresses, the gas costs for uploading this data to the blockchain would be prohibitive. Additionally, each update to the registry (adding or removing addresses) incurs gas costs and also needs to be continuous.

A more gas-efficient variation uses bit mapping within existing storage variables. For ERC-20 tokens, this can be implemented by using the most significant bits of the balance variable to store status flags:

\begin{lstlisting}[language=Solidity]
    mapping(address => uint256) internal balanceAndStatusFlags;
    
    // Bit positions for different flags
    uint256 constant EAI_FLAG_BIT = 255;      // Top bit for EAI status
    uint256 constant EXCHANGE_FLAG_BIT = 254; // Penultimate bit for exchange status
    
    function isEAI(address account) internal view returns (bool) {
        return (balanceAndStatusFlags[account] >> EAI_FLAG_BIT) & 1 == 1;
    }
    
    function isExchange(address account) internal view returns (bool) {
        return (balanceAndStatusFlags[account] >> EXCHANGE_FLAG_BIT) & 1 == 1;
    }
\end{lstlisting}

This approach is similar to the blocklisting implementation in the USDC contract \citep{CircleFinancial2023} and requires no additional storage variables, making it highly gas-efficient for both checks and updates. Since Ethereum's uint256 can store values up to $2^{256} - 1$, using the top few bits for boolean flags still leaves more than enough capacity for any realistic token balance ($2^{254}$ is approximately $10^{76}$, far exceeding the world's total wealth).

The bit mapping approach can be further optimized by implementing a dynamic EAI flagging mechanism that automatically updates the registry based on transaction patterns:

\begin{lstlisting}[language=Solidity]
    function transfer(address to, uint256 amount) 
        public override returns (bool) 
    {
        // Standard transfer logic
        // ...
        
        // If sender is an exchange, mark recipient as EAI
        if (isExchange(msg.sender)) {
            // Set the EAI flag bit for the recipient
            balanceAndStatusFlags[to] |= (1 << EAI_FLAG_BIT);
        }
        
        return true;
    }
\end{lstlisting}

This dynamic approach reduces the need for manual registry updates and aligns with the natural flow of funds from exchanges to user wallets. By automatically flagging addresses that receive funds directly from exchanges, the system can maintain an up-to-date EAI registry without requiring separate transactions for registry maintenance.

\subsubsection*{Off-Chain Registry with On-Chain Verification}

To address the high cost of onchain storage, an alternative approach maintains the EAI registry off-chain while using cryptographic signatures for onchain verification:

\begin{itemize}
    \item \textbf{Registry Storage:} The complete list of EAI addresses is maintained by a trusted entity off-chain
    \item \textbf{Transaction Process:} When initiating a transaction, users obtain a signed attestation that their address is in the EAI registry
    \item \textbf{Verification:} The smart contract verifies the signature to confirm EAI status
\end{itemize}

This approach dramatically reduces onchain storage costs but introduces new considerations. First, the method introduces the need for a central registry, as this method creates a reliance on an off-chain entity to maintain the registry and provide signatures. Additionally, there is transaction overhead to consider, since each transaction must include signature data, which increases calldata size. Finally, the approach results in limited transparency because the complete registry is not directly visible onchain.

The verification process typically requires checking signatures for both the sender and receiver in a transaction, adding computational overhead but eliminating the need for extensive onchain storage.

\subsubsection*{On-Chain Merkle Trees}

Merkle trees offer an elegant compromise between onchain and off-chain approaches \citep{buterin2022blockchain}. This implementation:

\begin{itemize}
    \item Stores only the Merkle root hash onchain (a single 32-byte value)
    \item Requires users to provide a Merkle proof with their transactions
    \item Verifies the proof onchain to confirm EAI status
\end{itemize}

The key advantage is that updating the registry requires changing only a single value (the Merkle root), regardless of how many addresses are added or removed. This makes large-scale updates extremely gas-efficient compared to the onchain registry approach. OpenZeppelin provides standard libraries for Merkle proof verification, making implementation straightforward. 

However, this approach has two notable drawbacks. First, there is increased transaction complexity, as users must obtain and include Merkle proofs with their transactions, typically requiring interaction with an off-chain service. Second, there are higher verification costs to consider, since proof verification requires O(log n) hash operations, where n is the size of the registry, which increases per-transaction gas costs.

\subsubsection*{Performance Comparison}

We implemented and tested each approach using a modified ERC-20 token contract with added EAI verification. Table \ref{table:nonlin} presents the gas costs for key operations across different implementation strategies.

\begin{table}[ht]
\caption{Transaction Cost Comparison} 
\centering 
\begin{tabular}{rcrrrr} 
\toprule
Whitelist Size & Method & isEAI gas & isEAI \$ & transfer gas & transfer \$\$ \\ 
\midrule
- & On-Chain & 612 & \$0.03 & 56,033 & \$2.69 \\
- & Off-Chain & 6,757 & \$0.32 & 65,112 & \$3.13 \\
500 & Merkle & 6,283 & \$0.30 & 70,273  & \$3.37\\
30,000 & Merkle & 8,214 & \$0.39 & 79,393  & \$3.81 \\
2,250,000 & Merkle & 10,135 & \$0.49 & 88,475 & \$4.25 \\
\bottomrule
\end{tabular}
\label{table:nonlin} 
\begin{flushleft}
\small{Note: Costs calculated using \$2,400 per ETH and 20 gwei gas price. Transfer transactions require two EAI checks: one for sender and one for receiver.}
\end{flushleft}
\end{table}

Our findings reveal that onchain registry verification is the most gas-efficient for individual transactions, but this advantage must be weighed against the high initial deployment and update costs. For the Merkle tree approach, verification costs increase logarithmically with registry size, but remain manageable even for very large registries.

The cost of adding new addresses to the registry varies dramatically across approaches:
\begin{itemize}
    \item \textbf{Off-Chain Registry:} No onchain cost (updates happen off-chain)
    \item \textbf{Merkle Tree:} Fixed cost of approximately 26,785 gas (\$2.16) regardless of how many addresses are added
    \item \textbf{On-Chain Registry:} Approximately \$2.30 per address, with practical limits of about 2 million addresses per transaction due to block gas limits
\end{itemize}

\subsection{Hybrid Approaches}

Given these tradeoffs, hybrid implementations may offer the best balance for practical applications. For example:

\begin{itemize}
    \item Store frequently accessed addresses (such as hot wallets of major exchanges) in an onchain registry for efficient verification
    \item Use Merkle proofs for less common addresses to reduce storage requirements
    \item Implement dynamic EAI flagging for addresses receiving funds directly from known exchanges
\end{itemize}

This tiered approach optimizes gas costs while maintaining the security benefits of EAI-based access control. High-volume users could even pay to "upgrade" their addresses from the Merkle-verified tier to the onchain registry to reduce their per-transaction costs.

\FloatBarrier

\section{Conclusion}

This paper has introduced the complementary concepts of Transaction Proximity and Easily Attainable Identities (EAIs) as a novel approach to identity and risk assessment on public blockchains. Our research demonstrates that a significant majority of legitimate blockchain activity occurs within close proximity to centralized exchanges, creating an opportunity for security enhancement without sacrificing the fundamental principles of decentralization and user privacy.

Our empirical analysis of the Ethereum blockchain revealed several key findings:

\begin{enumerate}
    \item \textbf{Prevalence of EAIs:} 56\% of USDC wallets with max lifetime balances exceeding \$10,000 are directly identifiable as EAIs, and 88\% are within one transaction hop of EAIs.
    \item \textbf{Transaction Traceability:} 91\% of USDC transactions exceeding \$2,000 involve at least one EAI wallet, indicating high traceability within the ecosystem.
    \item \textbf{Deterrence Potential:} 83\% of known exploiter addresses are not EAIs, with 21\% being more than five hops away from any EAI, suggesting that EAI-based access control could effectively deter malicious actors.
    \item \textbf{Implementation Feasibility:} We demonstrated three viable approaches for implementing EAI-based access control systems, each with different tradeoffs regarding gas costs, transparency, and decentralization.
\end{enumerate}

These findings have significant implications for blockchain security and the future development of DeFi. Transaction proximity analysis and EAI-based access validation represent a middle ground between the pseudo-anonymity that enables fraud and the strict KYC requirements that hinder adoption. By leveraging the natural structure of transaction networks, we can create security mechanisms that work with—rather than against—the architecture of blockchain systems.

It is important to emphasize that our approach does not restrict access to blockchain protocols or impose universal requirements. Instead, it provides valuable information that protocols and exchanges can use to implement their own validation or risk scoring systems based on their specific needs and risk tolerance. This approach respects the open nature of public blockchains while giving individual protocols the tools they need to make informed decisions about security and compliance.

The approach we propose offers several advantages over traditional methods:

\begin{itemize}
    \item \textbf{Preventative Rather Than Reactive:} Unlike blocklisting, which occurs after exploits, EAI-based controls can help prevent unauthorized access before damage occurs.
    \item \textbf{Privacy-Preserving:} Unlike universal KYC requirements, our approach does not require direct identification of all users, preserving privacy for the majority of participants.
    \item \textbf{Adoption-Friendly:} The system works with existing user behaviors and exchange relationships, requiring minimal changes to user experience.
    \item \textbf{Programmable Compliance:} Smart contract implementations enable automated, transparent enforcement without centralized intermediaries.
    \item \textbf{Flexible Implementation:} Protocols can choose how to utilize transaction proximity and EAI data based on their specific needs, from simple monitoring to sophisticated risk scoring models.
\end{itemize}

However, we acknowledge several limitations and challenges that require further research:

\begin{itemize}
    \item \textbf{Dynamic Nature of Transaction Graphs:} As blockchain usage evolves, the patterns of connectivity between wallets may change, requiring adaptive approaches to EAI identification.
    \item \textbf{Cross-Chain Compatibility:} Our analysis focused on Ethereum; extending this approach to other blockchains and enabling cross-chain compatibility presents additional challenges.
    \item \textbf{Governance and Maintenance:} Determining who maintains EAI lists and how updates are governed requires careful consideration of centralization risks.
    \item \textbf{Privacy Enhancements:} While more privacy-preserving than universal KYC, further research into zero-knowledge proofs and other privacy technologies could enhance our approach.
\end{itemize}

Future research directions include refining transaction proximity methodologies, developing more efficient onchain verification mechanisms, exploring integration with emerging privacy technologies, and conducting longitudinal studies on the effectiveness of EAI-based controls in preventing exploits over time.

As the DeFi ecosystem continues to mature, striking the right balance between security and privacy will be crucial for sustained adoption. Transaction proximity and EAI-based access validation offer a promising path forward—one that respects the core values of open blockchain while addressing the legitimate concerns of users, developers, and regulators. 

\clearpage

\bibliography{references}

\begin{thebibliography}{9}
\providecommand{\natexlab}[1]{#1}
\providecommand{\url}[1]{\texttt{#1}}
\expandafter\ifx\csname urlstyle\endcsname\relax
  \providecommand{\doi}[1]{doi: #1}\else
  \providecommand{\doi}{doi: \begingroup \urlstyle{rm}\Url}\fi

\bibitem[Brummer(2022)]{brummer2022disclosure}
Chris Brummer.
\newblock Disclosure, dapps and defi.
\newblock \emph{Stanford Journal of Blockchain Law and Policy}, March 2022.
\newblock URL \url{https://ssrn.com/abstract=4065143}.
\newblock Forthcoming.

\bibitem[Buterin et~al.(2022)Buterin, Hitzig, and Weyl]{buterin2022blockchain}
Vitalik Buterin, Zoë Hitzig, and E~Glen Weyl.
\newblock Blockchain resource pricing.
\newblock \emph{ACM Transactions on Economics and Computation}, 10\penalty0
  (4):\penalty0 1--37, 2022.

\bibitem[Buterin et~al.(2024)Buterin, Illum, Nadler, Sch{\"a}r, and
  Soleimani]{buterin2024privacy}
Vitalik Buterin, Jacob Illum, Matthias Nadler, Fabian Sch{\"a}r, and Ameen
  Soleimani.
\newblock Blockchain privacy and regulatory compliance: Towards a practical
  equilibrium.
\newblock \emph{Blockchain: Research and Applications}, 5\penalty0
  (1):\penalty0 100176, 2024.
\newblock \doi{10.1016/j.bcra.2023.100176}.

\bibitem[{Chainalysis}(2024)]{chainalysis_crypto_crime_2024}
{Chainalysis}.
\newblock The chainalysis 2024 crypto crime report.
\newblock Report, Chainalysis, 2024.
\newblock URL \url{https://go.chainalysis.com/crypto-crime-2024.html}.

\bibitem[Duffie et~al.(2025)Duffie, Olowookere, and Veneris]{duffie2025privacy}
Darrell Duffie, Odunayo Olowookere, and Andreas Veneris.
\newblock A note on privacy and compliance for stablecoins.
\newblock Technical report, Stanford University, York University, University of
  Toronto, March 2025.
\newblock Preliminary Draft.

\bibitem[Financial(2023)]{CircleFinancial2023}
Circle~Internet Financial.
\newblock Fiattokenv2\_2.sol, 2023.
\newblock URL
  \url{https://github.com/circlefin/stablecoin-evm/blob/master/contracts/v2/FiatTokenV2_2.sol}.

\bibitem[Gross et~al.(2022)Gross, Sedlmeir, and Seiter]{gross2022compliant}
Jonas Gross, Johannes Sedlmeir, and Simon Seiter.
\newblock How to design a compliant, privacy-preserving fiat stablecoin via
  zero-knowledge proofs.
\newblock Technical report, etonec GmbH, University of Luxembourg, Hauck
  Aufhäuser Lampe, 2022.
\newblock Technical Report.

\bibitem[Ron and Shamir(2013)]{ron2013quantitative}
Dorit Ron and Adi Shamir.
\newblock Quantitative analysis of the full bitcoin transaction graph.
\newblock In \emph{Financial Cryptography and Data Security}, pages 6--24,
  Berlin, Heidelberg, 2013. Springer.
\newblock Analysis conducted on Bitcoin blockchain data through May 13, 2012.

\bibitem[Taher et~al.(2024)Taher, Ameen, and Ahmed]{taher2024fraud}
Shimal~Sh. Taher, Siddeeq~Y. Ameen, and Jihan~A. Ahmed.
\newblock Advanced fraud detection in blockchain transactions: An ensemble
  learning and explainable ai approach.
\newblock \emph{Engineering, Technology \& Applied Science Research},
  14\penalty0 (1):\penalty0 12822--12830, 2024.
\newblock \doi{10.48084/etasr.6641}.

\end{thebibliography}
\newpage
\section{Appendix}
\subsection{Definition of USDC Wallet-to-Wallet Transaction}
A USDC wallet-to-wallet transaction is defined as a transfer of USDC initiated by calling the USDC smart contract directly (address 0xA0b86991c6218b36c1d19D4a2e9Eb0cE3606eB48 on the Ethereum network), as specified by the \texttt{to} field of the transaction. 

This represents the simplest form of USDC transfers, where wallet A transfer some amount of USDC to wallet B. More complex USDC transfers, such as swapping USDC for ETH on decentralized exchanges, are excluded.
\subsection{Code for Replications}
Github repository: \url{https://github.com/ziming494/easily_attainable_identities}

\end{document}